\documentclass[aps,prmaterials,reprint,superscriptaddress]{revtex4-2}
\usepackage{graphicx}
\usepackage[version=4]{mhchem}
\usepackage{amssymb}
\usepackage{bm}% bold math
\usepackage{epsfig}
\usepackage{graphicx}
\usepackage{amsmath}
\usepackage{color}
\usepackage{longtable}
\usepackage{inputenc}
\usepackage{tabularx}
\usepackage[normalem]{ulem} % delete this!
\usepackage{comment} %delete
\usepackage{tabularray}
\usepackage[margin=0.75in]{geometry} % Adjust the margins here (0.75in is an example)

\usepackage{xcolor}

\DeclareUnicodeCharacter{2212}{-}
\begin{document}
\title{The dipolar Aleppo lattice: Ground state ordering and ergodic dynamics in the absence of vertex frustration}

\author{Gopi Mahato}
\affiliation{Department of Physics, Baylor University,~One Bear Place, TX 76798, Waco, USA}
\author{Davis~Crater}
\affiliation{Department of Physics, Baylor University,~One Bear Place, TX 76798, Waco, USA}
\author{Christian~Hoyt}
\affiliation{Department of Physics, Baylor University,~One Bear Place, TX 76798, Waco, USA}
\author{Duncan Miertschin}
\affiliation{Department of Physics, Baylor University,~One Bear Place, TX 76798, Waco, USA}
\author{Balaram~Regmi}
\affiliation{Department of Physics, Baylor University,~One Bear Place, TX 76798, Waco, USA}
\author{Kevin~Hofhuis}
\affiliation{Paul Scherrer Institut, Forschungsstrasse 111, 5232 Villigen PSI, Switzerland}
\author{Scott Dhuey}
\affiliation{Molecular Foundry, Lawrence Berkeley National Laboratory, One Cyclotron Road, Berkeley, CA, 94720, USA}
\author{Barat Achinuq}
\affiliation{Lawrence Berkeley National Laboratory, One Cyclotron Road, Berkeley, California 94720, USA}
\author{Francesco Caravelli}
\affiliation{Theoretical Division (T4), Los Alamos National Laboratory, Los Alamos, NM, 87545, USA}
\author{Alan~Farhan}
\email{alan\_farhan@baylor.edu}
\affiliation{Department of Physics, Baylor University,~One Bear Place, TX 76798, Waco, USA}

\date{\today}

\begin{abstract}
We introduce the Aleppo spin ice geometry, another variation of decimated square ice patterns, which in contrast to similar systems previously studied, does not exhibit vertex frustration. Using synchrotron-based photoemission electron microscopy, we directly visualize low-energy states achieved after thermal annealing, in addition to temperature-dependent moment fluctuations. The results reveal the observation of ground state patterns and the absence of ergodicity-breaking dynamics. Our observations further confirm vertex frustration to be an important criterion for the emergence of ergodicity transitions. 

\end{abstract}

\maketitle

\section{Introduction}
Artificial spin ice systems, consisting of single-domain nanomagnets lithographically arranged onto two- and three-dimensional geometries, are ideal model systems for the direct visualization of the consequences of geometrical- and topological frustration~\cite{Skjaervo2020}. Prominent examples range from the observation of Coulomb phases with associated emergent magnetic monopoles~\cite{Perrin2016,Farhan2019,Ladak2010,Ostman2018}, triangular antiferromagnets~\cite{Farhan2020}, glassy networks~\cite{Saccone2020,Saccone2022}, phase transitions~\cite{Hofhuis2022,Schanilec2020,Leo2018}, programmable magnonics~\cite{Gliga2020, Iacocca2020,Jungfleisch2016} to a variety of emergent phenomena appearing in so-called vertex-frustrated artificial spin ice systems~\cite{Gilbert2014,Gilbert2015,Lao2018,Zhang2023,Saccone2023,Zhang2023,Stopfel2018,Saccone2019}. At the application frontier, artificial spin ice and its derivatives is being investigated for gate-based computing \cite{arava2018,gypens2018,caravellig} and machine learning  \cite{gunnar,gartside1,gartside2,vertex,amr1,amr2}. 

Vertex frustration is a form of magnetic frustration that emerges due to a topological competition of ordering patterns within four- and three-nanomagnet vertices in systems created by a controlled periodic decimation of artificial square ice patterns~\cite{Morrison2013}.  
Most recently, an alternative route towards vertex frustration in a stretched pentagonal lattice, void of any square ice vertices~\cite{Crater2024}.  

Within these vertex-frustrated systems, the most recent observation of ergodicity-breaking dynamics sparked a strong initial interest~\cite{Lao2018,Zhang2023,Saccone2023}. The interplay and competition of interactions and dynamics within and between four-nanomagnet and three-nanomagnet vertices was suggested to be the primary reason for the observed non-ergodic dynamics~\cite{Saccone2023}. However, questions loomed about whether this vertex-to-vertex frustration is the key ingredient in observing the mentioned ergodicity transitions. Therefore, exploring the validity of alternative system designs that share some geometrical similarities to vertex-frustrated systems is important, however, with one crucial difference: They shall lack this form of frustration. In other words, an artificial spin ice geometry needs to be introduced that is based on a decimated square ice, exhibiting a periodic mixture of four- and three-nanomagnet vertices, without the emergence of vertex frustration and showing clear preferences for long-range ordered patterns, particularly after employing established thermal annealing protocols on artificial spin ice~\cite{Farhan2013,Farhan2017,Farhan2019,Farhan2020}.

\begin{figure*}
	\centering
	\includegraphics[width=1.98\columnwidth]{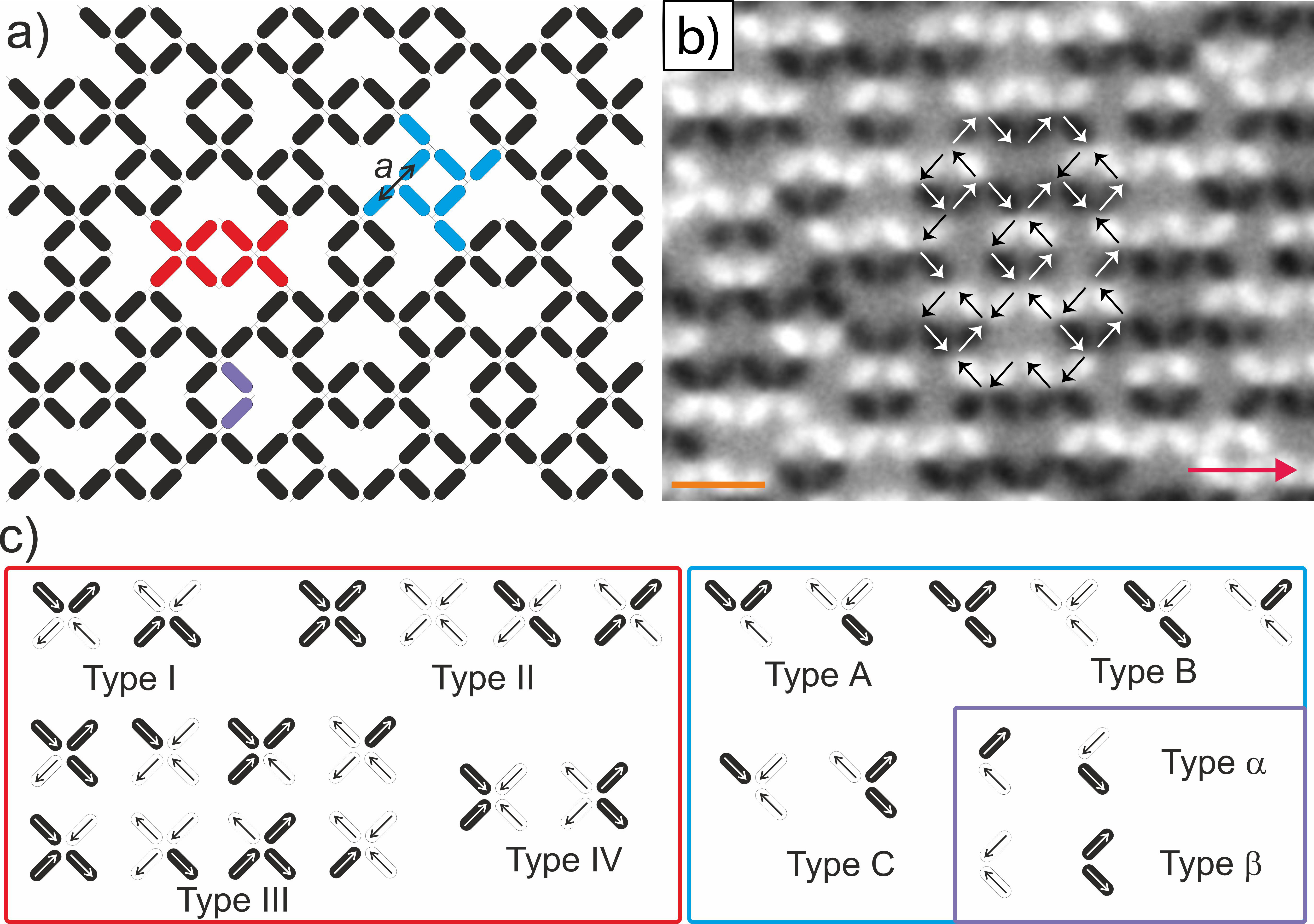}
	\caption{\label{fig:1} (a) Dipolar Aleppo lattice consisting of a mixture of four- and three-nanomagnet vertices (highlighted with red and blue islands, respectively). (b) XMCD image revealing magnetic configurations achieved after thermal annealing. Moments with a non-zero component pointing towards the incoming X-ray direction (red arrow) will appear dark, while moments in the opposite directions will appear bright. White and black arrows reveal a typical ground state formation. The orange bar corresponds to a length of around 1~$\mu$m. (c) Vertex types are achievable at four-, three- and two-nanomagnet vertices listed with increasing magnetostatic energy.}
\end{figure*} 

Here, we aim to address this point by introducing the dipolar Aleppo lattice (see Fig.\ref{fig:1}), a name derived due to similarities to historic tiles discovered in the city of Aleppo. It is a decimated square ice geometry where inter-nanomagnet dipolar coupling can be tuned by varying the lattice parameter $a$ (see Fig.\ref{fig:1}a). The main building block can be seen as a couple of square ice vertices (see red islands in Fig.\ref{fig:1}a). Then, a periodic repetition of horizontal and vertical arrangements of this building block forms the lattice. The result is a lattice exhibiting four-, three- and two-nanomagnet vertices. 

An alternative way to build up this lattice is a periodic arrangement of pinwheels consisting of four three-nanomagnet vertices (see blue islands in Fig.\ref{fig:1}a). Aside from these vertex types, a key feature in the Aleppo lattice is the L-shaped plaquettes interlinked via squares. Magnetic ordering patterns emerging within such shapes, for example clockwise- and anti-clockwise magnetic vortices, is a common feature in artificial spin ice systems~\cite{Farhan2013,Hofhuis2022,Saccone2023}. This is particularly evident when long-range ordered magnetic ground state patterns emerge. As the Aleppo lattice lacks vertex frustration, we expect low-energy states to be dominated by the presence of Type I and Type A vertices (see Fig.\ref{fig:1}c), at the four- and three-nanomagnet vertices, respectively. 

In the following, we will outline experimental methods with regard to sample fabrication and magnetic imaging of low-energy states achieved after thermal annealing and temperature-dependent fluctuations. We then discuss and analyze experimental results, followed by a summary and outlook to future endeavors, based on current findings. 

\section{Methods}
\subsection{Sample Fabrication \& Magnetic Imaging}

 Lift-off-assisted e-beam lithography was utilized to generate the dipolar Aleppo lattices studied here. A 1$\times$1~cm$^{2}$ Si (100) substrate is first spin coated with a 70~nm thick polymethylmethacrylate (PMMA, 950k). Then, a Raith EBPG 5000+ e-beam writer is used to place Aleppo patterns on top of the spin-coated substrate. After the exposed layer is developed, a 2.7~nm permalloy (Ni$_{80}$Fe$_{20}$) film is deposited, together with a 1.5~nm Aluminum capping layer, via thermal evaporation at a base pressure of $3\times10^{-7}$ Torr. Finally, a lift-off process in acetone was performed. The result is dipolar Aleppo lattices consisting of Ising-type nanomagnets with lengths $L =$ 400~nm, widths $W =$ 100~nm and thicknesses of 2.7~nm. Four sets of Aleppo lattices were patterned with four different lattice parameters $a$ (see Fig.\ref{fig:1}a) of 500~nm, 550~nm, 600~nm and 650~nm. 

 Magnetic imaging was performed at the cryogenic photoemission electron microscope (PEEM3) endstation of the Advanced Light Source~\cite{Doran2012}. We take advantage of the X-ray magnetic circular dichroism (XMCD) effect at the Fe L$_3$ edge~\cite{Stohr1993}. An XMCD image is obtained by pixel-wise division of images recorded with circular right-polarized X-rays and images recorded with circular left-polarized X-rays. The exposure time at each polarization was set at 1.5 seconds, while the beamline requires around four seconds to switch polarization. In other words, we are then able to obtain XMCD image sequences with a temporal resolution of around 7 seconds per frame. The chosen dimensions of the patterned nanomagnets ensure accessibility to thermally-driven moment reorientations at an accessible temperature range below 300~K~\cite{Farhan2019,Saccone2023}. The blocking temperature $T_{B}$ for the Aleppo patterned lattices, defined as the temperature at which fluctuations begin to emerge on the time scale of several seconds, was determined to be 210~K. 

\begin{figure*}
	\centering
	\includegraphics[width=1.98\columnwidth]{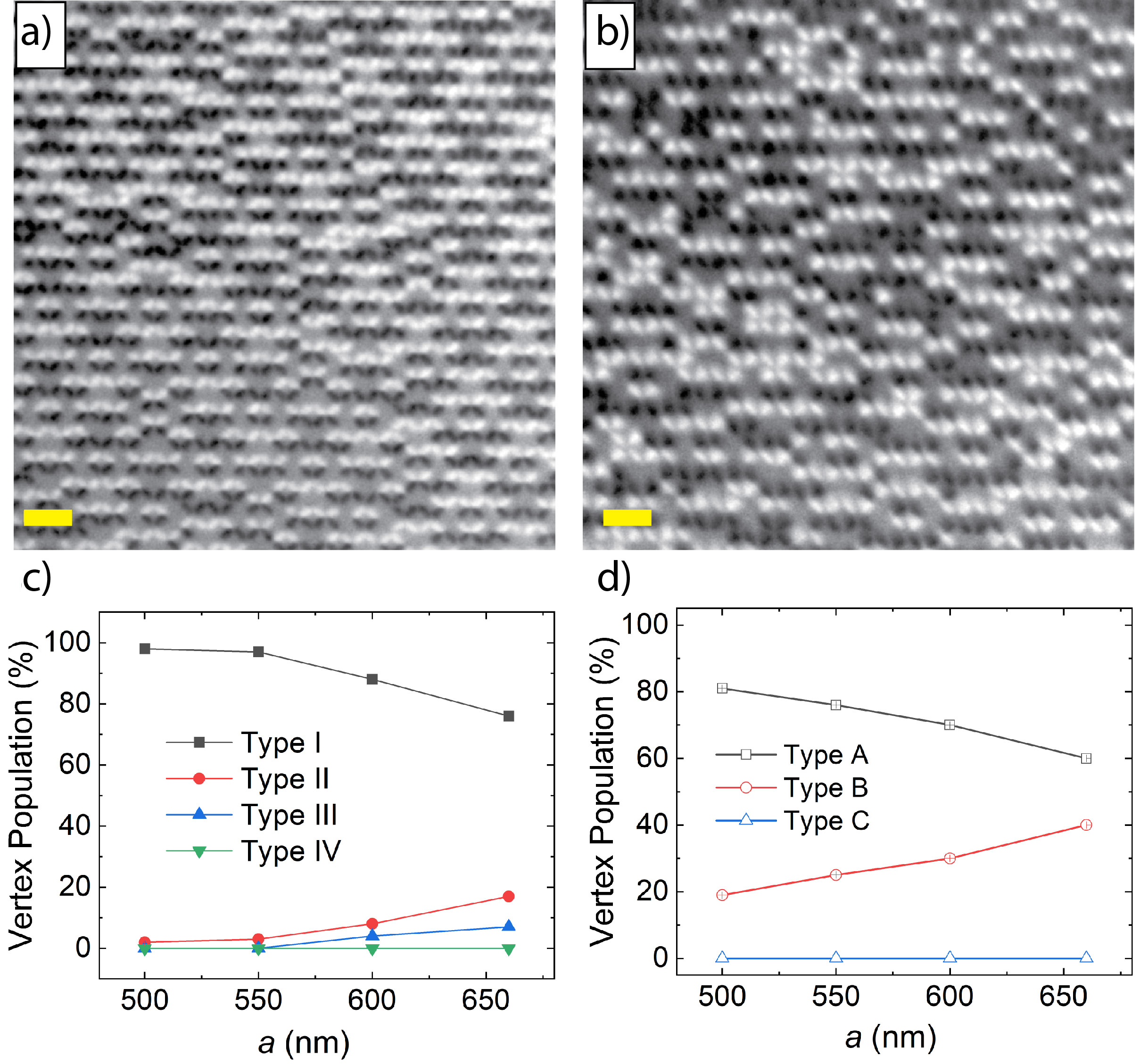}
	\caption{\label{fig:2} XMCD images of low-energy states achieved in the dipolar Aleppo lattices with lattice parameters (a) $a =$ 500~nm and (b) $a =$ 660~nm,  after thermal annealing. (c) Average Vertex types achieved at four-nanomagnet vertices, after thermal annealing, plotted as a function of the employed lattice parameter. (d) Vertex type statistics at three-nanomagnet vertices, following the same annealing procedures.}
\end{figure*} 

\section{Results}
\subsection{Thermal annealing and ground state ordering}
Following the fabrication steps mentioned above, the sample was kept at room temperature for more than three weeks, inside a vacuum box. Then it was transferred to PEEM3 and cooled down to 150~K. This allows for frozen magnetic moment configurations to be imaged after the three-weeks-long annealing at room temperature (see examples in Fig~\ref{fig:2}a, b). The low-energy states achieved after annealing show a clear dominance of Type I vertex types at four-nanomagnet vertices and Type A vertex types at three-nanomagnet vertices. These vertex types appear to emerge compatibly, leading to long-range ordered patterns, especially for smaller lattice parameters and stronger dipolar coupling between nanomagnets (see Fig.~\ref{fig:2}a, c, and d). With an increasing lattice parameter $a$, we see the expected decrease in the Type I and Type A vertex population and the decline in ordered patterns (see Fig.~\ref{fig:2}b, c, and d). 
Typical ground state formations are highlighted with black and white arrows on top of Fig.~\ref{fig:1}b. Looking at such formations, we observe that it does indeed fully support the combination of Type I and Type A to form long-range ordered tiles of such a cell. Altogether, it confirms the absence of vertex frustration in the Aleppo lattice, making it a perfect system for a direct comparison to geometrically similar vertex-frustrated systems~\cite{Gilbert2014,Gilbert2015,Saccone2023,Zhang2023}. To conduct such a comparison, direct visualization of temperature-dependent moment fluctuations is essential, to directly explore how dynamics at four- and three-nanomagnet vertex sites influence one another and how robust long-range order remains under the influence of thermal disturbances. Furthermore, it allows direct extraction of the temperature dependence of the stress metric ($\Omega$)~\cite{Saccone2023,Mountain1993}.

\subsection{Temperature-dependent fluctuations: Vertex populations, spin-spin correlations, stress metric and ergodicity}

\begin{figure}
	\centering
	\includegraphics[width=0.95\columnwidth]{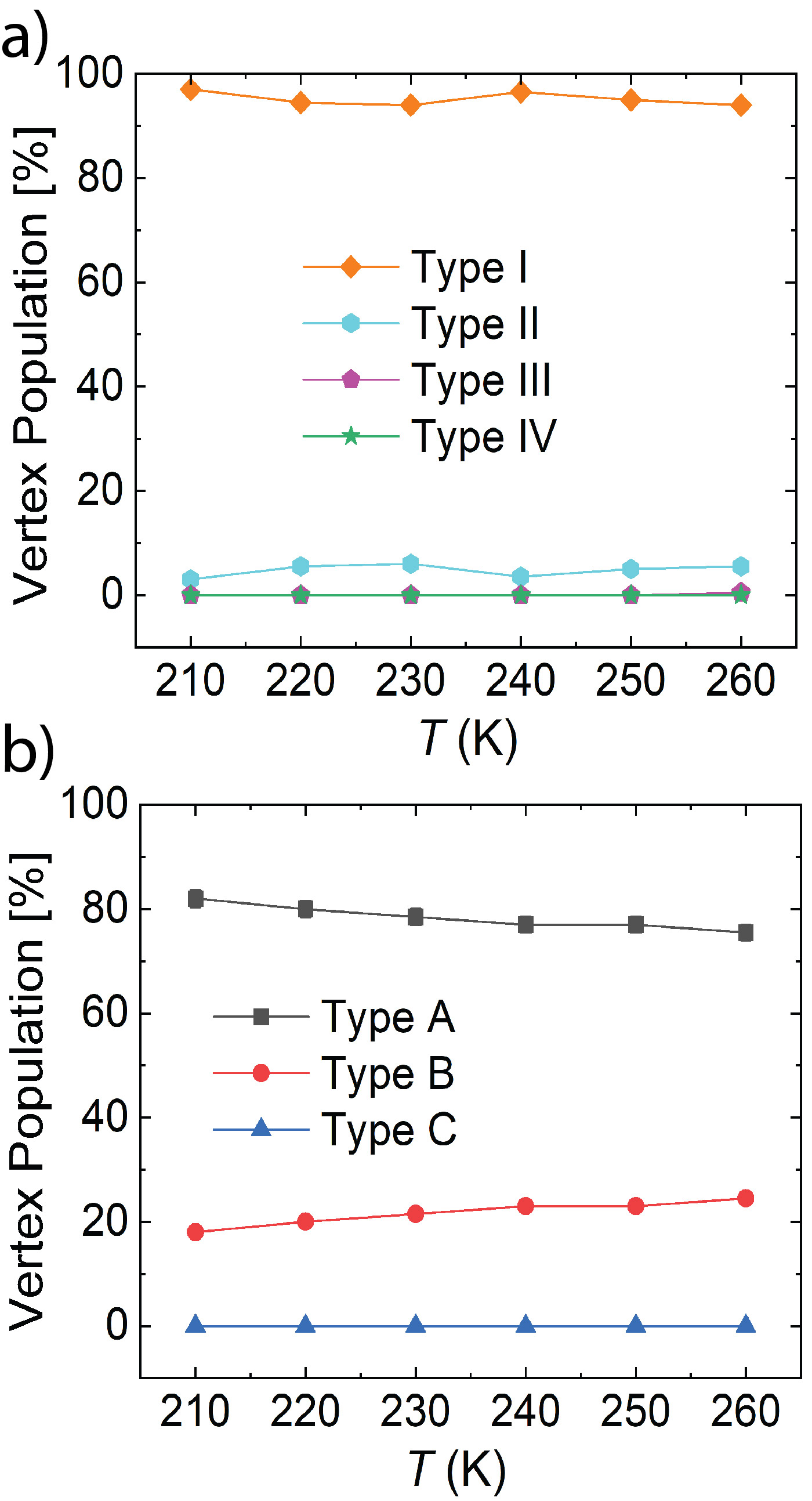}
	\caption{\label{fig:3} (a) Four-nanomagnet vertex type populations plotted as a function temperature. (b) Populations of three-nanomagnet vertex types plotted as a function of temperature.}
\end{figure}

Following the determination of the blocking temperature to be around 200~K, temperature-dependent XMCD imaging was started, recording 70-100 XMCD at six different temperatures, starting from 210~K up to 260~K (see examples of Supplementary Movie 1 and Supplementary Movie 2~\cite{SM}). Focusing on an Aleppo lattice with a lattice parameter $a =$ 500~nm, we see Type I vertex populations dominate and remain robust against increasing temperature (see Fig.~\ref{fig:3}a), while Type A vertex populations show a modest decrease with increasing temperature, mirrored by a corresponding increase in Type B population (see Fig.~\ref{fig:3}b). Overall, these observations confirm the robustness of the observed long-range ordered ground state and the total absence of vertex frustration in the Aleppo lattice. 

\begin{figure}
	\centering
	\includegraphics[width=0.95\columnwidth]{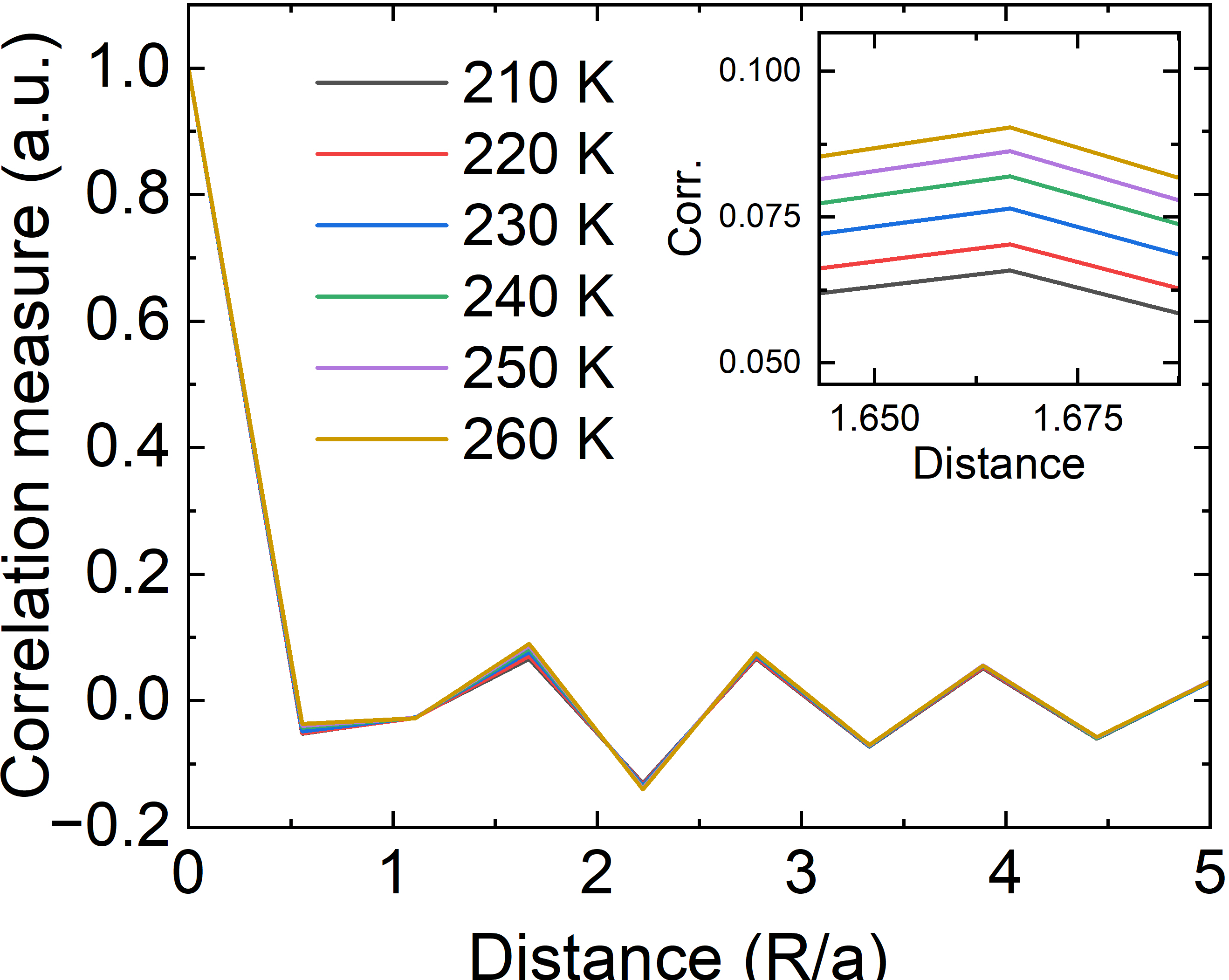}
	\caption{\label{fig:4} Spin-spin correlations plotted as a function of distances normalized against the lattice parameter $a =$ 500~nm at various temperatures.}
\end{figure}

\begin{figure}
	\centering
	\includegraphics[width=0.95\columnwidth]{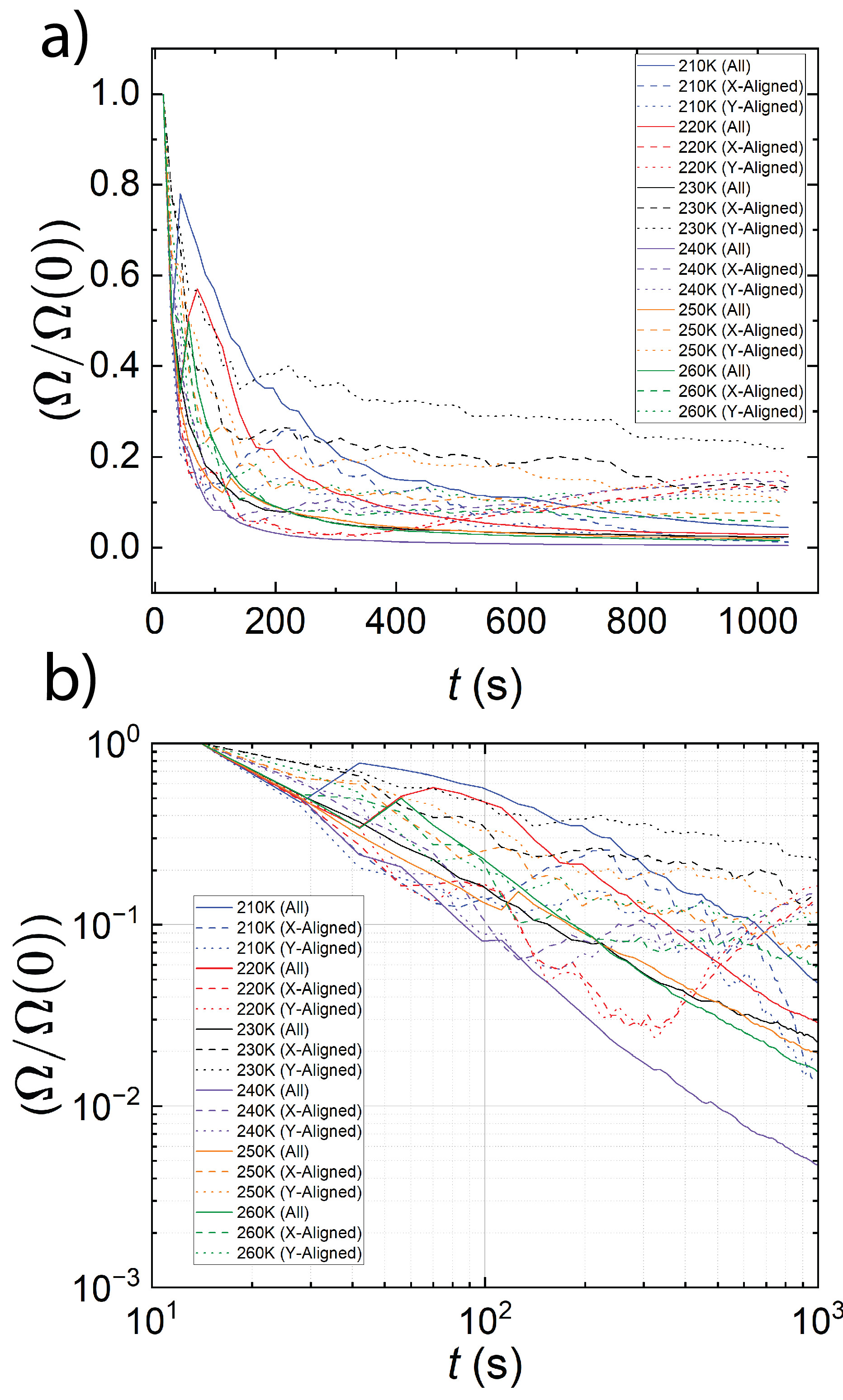}
    \caption{\label{fig:5} TM stress metric decay as a function of time in linear (a) and logarithmic scale (b). Dashed and and dotted lines represent behavior of magnetic moments along the x- and y-direction, respectively. The solid lines represent a combination of all involved magnetic moments.}
\end{figure}

\begin{figure}
	\centering
	\includegraphics[width=0.95\columnwidth]{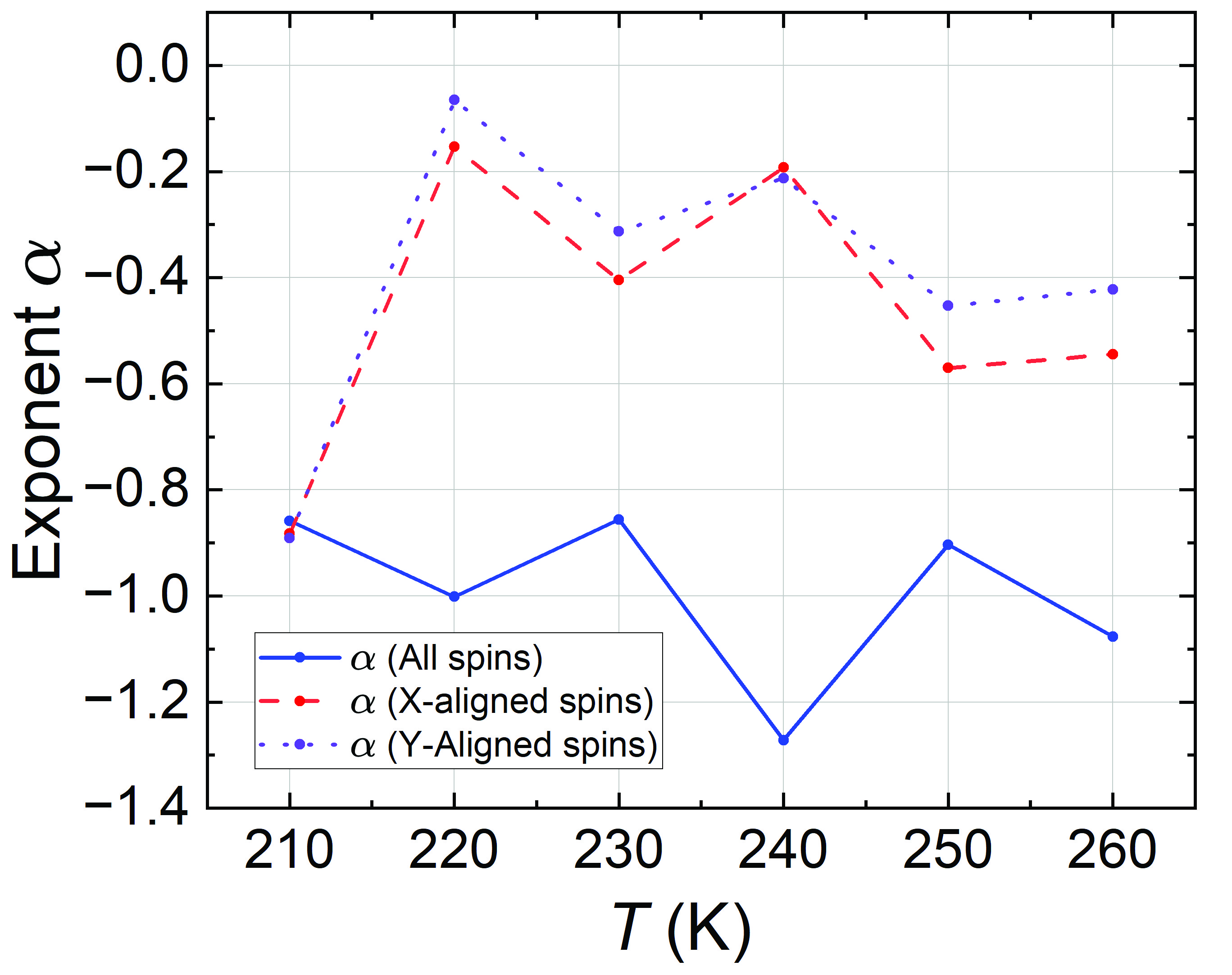}
    \caption{\label{fig:6} Exponent $\alpha$ as a function of temperature, fit for $\Omega(t)\approx t^\alpha$.}
\end{figure}

To shed further light on the temperature dependence of moment configurations in the dipolar Aleppo lattice, we extracted spin-spin correlation measures~\cite{Gilbert2014,Saccone2019,Crater2024}. To determine the correlation measure $C$ of a given magnetic moment with neighboring magnetic moments as a function of distance, the dipolar interactions between spins falling within an increasing radius are considered. Pairs of magnetic moments that minimize dipolar interactions are assigned a correlation measure of +1, while those that maximize dipolar interaction will be assigned a correlation value of -1.  The average correlation measures are then plotted as a function of distance $R$ normalized against the lattice parameter $a$ for all temperatures (see Fig.~\ref{fig:4}). Looking at Fig.\ref{fig:4}, we see spin correlation measures hopping periodically from negative to positive correlation measures with increasing distance. This not only confirms the long-range ordered nature of configurations achieved in the dipolar Aleppo lattice, but also highlights that the Aleppo lattice features an antiferromagnetic type of ordering, in analogy to two-dimensional artificial square ice~\cite{Alan2013}. 

The inset in Fig.~\ref{fig:4} shows a zoomed-in portion of the plotted curves revealing an increase in correlation measures with increasing temperatures at that distance.

\subsection{Ergodic behavior of the Aleppo lattice}

Following this confirmation, we analyze the temperature-dependent moment fluctuations in the Aleppo lattice using the Thirumalai-Mountain (TM) fluctuating metric~\cite{Saccone2023, Baccetti2024, Mountain1993}. This metric, denoted as $\Omega_G(t_k)$, provides a measure of configurational relaxation and fluctuation dynamics over time. Defined at a given time $t_k$ as:
\begin{equation}
    \Omega_G(t_k) = \frac{1}{N} \sum_{j=1}^N \left[ g_j(t_k) - \langle g(t_k) \rangle \right]^2,
\end{equation}
it quantifies the variance of the time-averaged components $g_j(t_k)$, where:
\begin{equation}
    g_j(t_k) = \frac{1}{k} \sum_{i=0}^{k} g_j(t_i),
\end{equation}
represents the time-averaged behavior of the $j$-th degree of freedom, and
\begin{equation}
    \langle g(t_k) \rangle = \frac{1}{N} \sum_{j=1}^N g_j(t_k),
\end{equation}
is the instantaneous ensemble average across all $N$ components \cite{suzen}.

To capture directional relaxation dynamics, staggered TM metrics $\Omega_{G,x}(t_k)$ and $\Omega_{G,y}(t_k)$ were introduced, isolating the contributions of components predominantly aligned along the $x$- and $y$-directions, respectively. These are calculated similarly to $\Omega_G(t_k)$, but the summation is restricted to subsets of components aligned with the corresponding directions, determined by angular thresholds.

Figure~\ref{fig:5} illustrates the evolution of $\Omega_G(t_k)$, $\Omega_{G,x}(t_k)$, and $\Omega_{G,y}(t_k)$ over time for temperatures ranging from 210~K to 260~K. A consistent $\Omega_G(t_k)$ decline is observed across all temperatures, indicating ergodic convergence as the system approaches equilibrium. The staggered metrics, $\Omega_{G,x}(t_k)$ and $\Omega_{G,y}(t_k)$, follow similar trends but differ quantitatively, highlighting anisotropic relaxation along the lattice's principal axes.

The rate of relaxation was quantified by fitting $\Omega_G(t_k)$, $\Omega_{G,x}(t_k)$, and $\Omega_{G,y}(t_k)$ to a power-law decay:
\begin{equation}
    \Omega_G(t_k) \sim t_k^{-\alpha},
\end{equation}
where $\alpha$ is the relaxation exponent. As shown in Fig.~\ref{fig:6}, $\alpha$ for $\Omega_G(t_k)$ decreases with increasing temperature, consistent with slower relaxation dynamics due to enhanced thermal fluctuations disrupting long-range correlations. The staggered metrics, $\Omega_{G,x}(t_k)$ and $\Omega_{G,y}(t_k)$, exhibit distinct temperature dependencies for $\alpha$, highlighting the directional nature of the relaxation process.

As demonstrated in this study, the Aleppo lattice lacks the vertex frustration typical of many artificial spin ice systems, which often leads to non-ergodic dynamics. We hypothesize that the absence of frustration underpins the observed ergodic and diffusive relaxation behavior, contrasting with the dynamics reported for the Apamea lattice \cite{Saccone2023}. The robust relaxation dynamics across all temperatures reinforce this conclusion, reflecting the diffusive equilibration characteristic of systems free of topological constraints on their degrees of freedom.

\section{Conclusions}

In this study, we investigated the dipolar Aleppo lattice, a novel artificial spin ice geometry lacking vertex frustration, and explored its ground state configurations and temperature-dependent dynamics. Using XMCD imaging, we confirmed the emergence of robust long-range ordered patterns dominated by Type I and Type A vertices at four- and three-nanomagnet sites, respectively. These results demonstrate the absence of vertex frustration in the Aleppo lattice, making it an ideal system for direct comparison to geometrically similar, vertex-frustrated spin ice systems.

Through temperature-dependent measurements, we characterized the relaxation dynamics of the system using the Thirumalai-Mountain stress metric, $\Omega(t)$, along with its staggered directional counterparts, $\Omega_x(t)$ and $\Omega_y(t)$, which isolate contributions from spins predominantly aligned along the lattice's principal axes. The results revealed consistent relaxation across all studied temperatures, with diffusive equilibration behavior reflected in the steady decline of $\Omega(t)$ over time. Fitting the stress metrics to a power-law decay enabled the extraction of the relaxation exponent $\alpha$, which was found to exhibit a systematic temperature dependence. Specifically, the exponent $\alpha$ decreased with increasing temperature, indicating slower relaxation dynamics driven by enhanced thermal fluctuations.

The staggered metrics, $\Omega_x(t)$ and $\Omega_y(t)$, exhibited similar trends but differed quantitatively, highlighting anisotropic relaxation dynamics in the Aleppo lattice. These findings underscore the directional dependence of moment reorientations and provide further insights into the system's thermal response.

We hypothesize that the lack of vertex frustration in the Aleppo lattice is a key factor underpinning the observed diffusive and ergodic relaxation dynamics. This contrasts with vertex-frustrated systems, where topological constraints often lead to non-ergodic behavior. The robust relaxation dynamics across all studied temperatures confirm that the Aleppo lattice equilibrates diffusively, offering a unique platform for studying relaxation processes in artificial spin ice systems free of frustration-induced constraints. These insights pave the way for further exploration of lattice designs, and for further understanding the dynamics of artificial spin systems.

\section{Acknowledgements}
The authors would like to thank Dr.~Michael Saccone for his support. This research used resources of the Advanced Light Source and the Molecular Foundry, which are DOE Office of Science User Facilities under contract no. DE-AC02-05CH11231. FC's work was conducted under the auspices of the National Nuclear Security Administration of the United States Department of Energy at Los Alamos National Laboratory (LANL) under Contract No. DE-AC52-06NA25396, and supported financially via DOE LDRD grant 20240245ER.

\end{document}